\documentclass[11pt,a4paper]{article}

\usepackage{jheppub_kim}

\usepackage{subfigure,amsmath,amssymb ,amsfonts, latexsym}

\usepackage[notcite,color,final]{showkeys}
\definecolor{refkey}{gray}{.25}
\definecolor{labelkey}{gray}{.25}

\def\be{\begin{equation}}
\def\ee{\end{equation}}
\def\ba{\begin{eqnarray}}
\def\ea{\end{eqnarray}}

\def\be{\begin{equation}}
\def\ee{\end{equation}}
\def\bea{\begin{eqnarray}}
\def\eea{\end{eqnarray}}

\begin{document}

 \title{Density Perturbation Growth in Teleparallel Cosmology}

\author[a,b,c]{Chao-Qiang Geng}
\author[b]{Yi-Peng Wu}

\affiliation[a]{College of Mathematics \& Physics, Chongqing University of Posts \& Telecommunications, Chongqing, 400065, China}

\affiliation[b]{Department of Physics, National Tsing Hua University,
Hsinchu, Taiwan 300}

\affiliation[c]{National Center for Theoretical Sciences, Hsinchu,
Taiwan 300}

\emailAdd{geng@phys.nthu.edu.tw}
\emailAdd{s9822508@m98.nthu.edu.tw}



\abstract{
We study the cosmological perturbations in teleparallel dark energy models 
in which there is a dynamical scalar field with a non-minimal coupling to gravity.
We find that  the  propagating degrees of freedom are the same as in quintessence cosmology 
despite that variables of the perturbed vierbein field are greater than those in metric theories.
We numerically show some evident discrepancy from general
relativity in the evolutions of the perturbations on all scales of the universe.
We also demonstrate that the gravitational interactions are enhanced during the unique tracker evolutions in 
these models.}

\maketitle

\section{Introduction}

The dark energy mystery has received much attention in
the context beyond simply embedding the
cosmological constant in general relativity (GR).
To avoid a gigantic fine-tuning issue when the
cosmological constant is attributed to the vacuum
energy density, a canonical scalar field is introduced
with the dynamical behavior known
as quintessence~\cite{quintessence,quintessence1,
quintessence2,quintessence3,quintessence4,quintessence5}.
Further modification of quintessence has been
considered by allowing a non-minimal coupling
between the scalar field and gravity~\cite{nonminimal,nonminimal1,
nonminimal2,nonminimal3,nonminimal4,
nonminimal6,nonminimal7,nonminimal8,nonminimal9} as well as
 either using the ``wrong'' sign phantom field~\cite{phant,phant1,phant2,phant3,phant4,phant5},
or  combining a non-canonical extension
such as the scalar-tensor theories~\cite{st,Maedabook}.

Recently, the ``teleparallel'' description of
GR~\cite{ein28,Hayashi79} reveals
an interesting aspect of gravity to explain the
dark energy origin.
The late-time cosmic acceleration has been seen from the 
generalized action of teleparallel equivalence of 
GR (TEGR)~\cite{Maluf:1994ji},
%
named as $f(T)$ gravity~\cite{f(T)} where $T$ 
denotes the gravity Lagrangian of TEGR.
Although  $f(T)$ gravity
shows a clear
reference to  $f(R)$ gravity,
it provides
no higher than second-order field equations, and
no trivial Einstein frame avalible through conformal 
transformations~\cite{Yang:2010ji}. 
On the other hand, teleparallel gravity incorporating 
with a dynamical scalar field describes a different
 scenario to drive the accelerated expansion of the Universe.
This class of teleparallel dark energy 
models~\cite{TDE,Do TDE,PS TDE}
lies in the framework of TEGR where the simplest
quintessence model~\cite{quintessence} is 
recovered if the scalar field is minimally coupled
to gravity. Given that the torsion scalar
$T$ differs from the Lagrangian density of GR
by a total derivative, the presence of a non-minimal
coupling between the scalar field and gravity leads to interesting cosmological 
behaviors~\cite{quintessence,nonminimal,phant}, 
while teleparallel cosmology shows 
compelling results with current observations~\cite{ob TDE}.

However, it is noteworthy that modified teleparallel gravity theories,
including both $f(T)$ gravity and teleparallel dark energy models,
do not respect to the Lorentz invariance in the local
tangent frame. The Lorentz violation in the teleparallel formalism 
beyond TEGR inevitably introduces extra degrees of
freedom, which are unfamiliar in the framework of GR.
This fact is manifestly provided by the field equations, 
which address the dynamics of all  16 components of the
vierbein field \cite{LV, P within T}.
Under the homogeneous and isotropic principles of 
the background cosmology, those extra degrees of freedom merely
involve at perturbation level, but  have significant 
effects on the structure formulation  in the Universe~\cite{LSS}.

In the present work, we study the cosmological perturbations
in teleparallel dark energy models via the perturbed vierbein
field, which contains some variables with 
no reference in metric perturbations \cite{MDP in MTT}.
Nevertheless, in the scalar perturbations, we find that only
one new degree of freedom involves at linear level, and
 we obtain one implicit constraint equation to the scalar
mode variables from the asymmetric part of the 
field equations. As a result, the new scalar mode is 
resolved with a mere algebraic relation to other
scalar modes, and thus the propagating 
degrees of freedom in teleparallel cosmology do
not increase.
In the potential-less case, we numerically examine the evolution of the cosmological
perturbations and show that the anisotropic between the
gravitational potentials, addressing the effects
of the non-dynamical new degree of freedom, 
is significant on all scales of the universe.

We also proceed the density perturbations
in sub-horizon scales to study the growth behaviors
of some specific teleparallel dark energy models.
For the simplest tracker solution corresponding to the 
non-minimal gravity-field coupling~\cite{Gu}
(hereafter we denote as ``NGF tracker''), we compare
the growth rate from the effective gravitational constant
to the GR case  under an identical evolution background.
We find that the NGF tracker always dominates,
even in the presence of potentials of the scalar field,
if the coupling between gravity and the scalar field is large to the
order of unity. In the broader case  with a smaller 
non-minimal coupling, the cosmic acceleration can be
driven by the potential or the NGF tracker, depending on
the initial composite differences of dark energy.
The resulting density perturbation growth indicates that
gravitational interactions are stronger if the evolution follows
the NGF tracker rather than 
the usual potential-driven cosmic acceleration case.

This paper is organized as follows.
In Sec. II, we review teleparallel dark energy theories.
In Sec. III, we demonstrate the perturbation equations and
 apply the results to study the matter growth in Sec. IV.
Finally, conclusions are given in Sec. V.

\section{Teleparallel Dark Energy}\label{Tele DE}

The teleparallel dark energy model is 
given by~\cite{TDE}:
\begin{equation}
S=\int\,d^{4}x e\Bigg[\frac{T}{2\kappa^{2}}
+ \frac{1}{2} \Big(\partial_{\mu}\phi\partial^{\mu}\phi+\xi
T\phi^{2}\Big) - V(\phi)+\mathcal{L}_m\Bigg], \label{action}
\end{equation}
where $e\equiv\mbox{det}(e^A_\mu)=\sqrt{-g}$
with  $\kappa^{2}=8\pi G$, 
$\xi$ is the non-minimal coupling parameter, and
\begin{equation} 
T= \frac{1}{4}T_{\rho}^{\,\,\mu\nu}T^{\rho}_{\,\,\mu\nu}-
    \frac{1}{2}T^{\mu\nu}_{\,\,\,\,\,\,\,\,\rho} T^{\rho}_{\,\,\mu\nu}-
               T^{\rho}_{\,\,\rho\mu}T^{\nu\,\,\mu}_{\,\,\nu}
\end{equation}
is the torsion scalar of TEGR specially constructed with
the quadratics of the torsion tensor~\cite{Maluf:1994ji}:
\begin{equation}  \label{torsion}
{T}^\lambda_{\:\mu\nu}=\Gamma^\lambda_{
\nu\mu}-%
\Gamma^\lambda_{\mu\nu}
=e^\lambda_A\:(\partial_\mu
e^A_\nu-\partial_\nu e^A_\mu).
\end{equation}
It is noticeable that the connection
$\Gamma^\lambda_{\mu\nu}=e^\lambda_A\partial_\nu e^A_\mu$
used in TEGR  not only describes a curvature-less
geometry but also guarantees the independent parallel
transformation among the four unit vectors
${\mathbf{e}_A(x^\mu)}$, 
which form an orthonormal basis of the tangent space:
$\mathbf{e}_A\cdot%
\mathbf{e}_B=\eta_{AB}$, 
where $\eta_{AB}=diag (1,-1,-1,-1)$
\cite{Hayashi79}.
These vectors, $\mathbf{e}_A$, are commonly addressed by the
components $e_A^\mu$ in the coordinate basis, 
$i.e.$ $\mathbf{e}_A = e^\mu_A\partial_\mu$~
\footnote{We use the notations as
follows: Greek indices $\mu, \nu,$... 
and capital Latin indices $A, B, $...
run over all coordinate and tangent
space-time 0, 1, 2, 3, while lower case Latin indices 
(from the middle of the
alphabet) $i, j,...$ and lower case Latin indices
(from the beginning of the alphabet) $a,b, $...  
run over spatial and tangent
space coordinates 1, 2, 3, respectively.},
 while the metric tensor is given by
the dual vierbein as 
$g_{\mu\nu}=\eta_{AB}\, e^A_\mu \, e^B_\nu $.

The field equations 
are obtained from the variation with respect to the 
vierbein, $e^A_\mu$, which is the dynamical field of 
teleparallel gravity, given by
\begin{eqnarray} \label{eom}
\left(\frac{1}{\kappa^2}+ \xi\phi^2 \right)G_{A}^\nu
 -e_{A}^{\nu}\left[\frac{1}{2}\partial_\mu\phi\partial^\mu\phi-V(\phi)\right]
 +e_A^\mu\partial^\nu\phi\partial_\mu\phi\ \ \nonumber\\
 + 4\xi S_{A}{}^{\lambda\nu}\phi\left(\partial_\lambda\phi\right)
=\Theta_{A}{}^{\nu},~~~~
\end{eqnarray}
where 
$\Theta_{A}{}^{\nu}\equiv e^{-1}\delta\mathcal{L}_m 
/\delta e^A_\nu$ denotes the energy-momentum tensor 
of the matter source, and
\begin{equation} 
\nonumber
G_{A}^\nu=
2e^{-1}\partial_{\mu}(ee_{A}^{\rho}S_{\rho}{}^{\mu\nu})
 -2e_{A}^{\lambda}T^{\rho}{}_{\mu\lambda}S_{\rho}{}^{\nu\mu}
 -\frac{1}{2}e_{A}^{\nu}T
\end{equation}
is nothing but the equivalent geometrical structure of the
Einstein tensor with $G_{\mu\nu}=e^A_\mu G_{A\nu}$,
where
\begin{equation} 
\nonumber
S_\rho^{\:\:\:\mu\nu}=
\frac{1}{2}\Big(K^{\mu\nu}_{\:\:\:\:\rho}
+\delta^\mu_\rho
\:T^{\alpha\nu}_{\:\:\:\:\alpha}-\delta^\nu_\rho\:
T^{\alpha\mu}_{\:\:\:\:\alpha}\Big)
\end{equation}
with
$K^{\mu\nu}_{\:\:\:\:\rho}=-\frac{1}{2}\Big(T^{\mu\nu}_{
\:\:\:\:\rho}
-T^{\nu\mu}_{\:\:\:\:\rho}-T_{\rho}^{\:\:\:\:\mu\nu}\Big)$.
We can see that in the minimal limit ($\xi=0$), the field equation
(\ref{eom}) is reduced to the simplest quintessence model.

Torsion in (\ref{torsion})
is defined by a Lorentz violated formalism for the tangent frame~\cite{Maluf:1994ji}, 
which provides a technically 
simpler approach to acquire results
of the cosmological interest~\cite{Blagojevic}. 
The lack of Lorentz symmetry
in teleparallelism results in  extra degrees of
freedom~\cite{LV} as seen from (\ref{eom}), which  in fact exhibits the
equation of motion of all the 16 degrees of freedom of
the vierbein field. To be more precise, 
given that the  energy-momentum tensor
$\Theta_{\mu\nu}$ and the Einstein tensor $G_{\mu\nu}$ are
both symmetric, the anti-symmetrization between  indices
${\small A}$ and $\nu$ in (\ref{eom})
leads to a non-trivial constraint
\begin{equation} \label{constraint}
4\xi\phi\left(g^{\mu\alpha}S_{\mu}{}^{\lambda\beta}-
              g^{\nu\beta} S_{\nu}{}^{\lambda\alpha}\right)\partial_\lambda\phi
              =0,
\end{equation}
which forms a system of 6 equations in addition to
the usual 10 equations from the symmetric part
of the field equations.
Incidentally, this constraint vanishes if $\xi=0$
 so that TEGR shares the same number of
dynamical degrees of freedom as GR.

Nevertheless, we will see in the later discussion that
only one extra scalar mode beyond the metric perturbation
variables will be involved in the linear
perturbations of teleparallel dark energy theories.
However,
 this new degree of freedom is not dynamically
independent due to the constraint in~(\ref{constraint}).

The flat Friedmann-Robertson-Walker (FRW) background metric
\begin{equation}
\nonumber
ds^2= dt^2-a^2(t)\delta_{ij} dx^i dx^j
\end{equation}
 is conventionally
resolved by the background vierbein choice
\begin{equation}  \label{FRWvierbeins}
e_{\mu}^A=\mathrm{diag}(1,a,a,a),
\end{equation}
where $a(t)$ is the scale factor.
The Friedmann equations are given by
\begin{eqnarray}
\label{FR1}
&&
H^{2}=\frac{\kappa^2}{3}\Big(\rho_{\phi}+\rho_{m}\Big),
\\
\label{FR2}
&&
\dot{H}=-\frac{\kappa^2}{2}\Big(\rho_{\phi}+p_{\phi}+\rho_{m}+p_{m}
\Big),~~~~
\end{eqnarray}
where 
\begin{eqnarray}
\label{telerho}
 &&\rho_{\phi}=  \frac{1}{2}\dot{\phi}^{2} + V(\phi)
-  3\xi H^{2}\phi^{2},\\
 &&p_{\phi}=  \frac{1}{2}\dot{\phi}^{2} - V(\phi) +   4 \xi
H \phi\dot{\phi}
 + \xi\left(3H^2+2\dot{H}\right)\phi^2\,,
 \label{telep}
\end{eqnarray}
with
$H=\dot{a}/a$  the Hubble parameter.
The variation of the action with 
respect to the scalar field yields
\begin{equation} \label{eom phi}
\square\phi-\xi T\phi+V_\phi=0,
\end{equation}
where $V_\phi\equiv dV/d\phi$.
Consequently, the background equation motion of $\phi$
is given by
\begin{equation} 
\ddot{\phi}+3H\dot{\phi}+6\xi H^2\phi+ V_\phi=0.
\label{fieldevol2}
\end{equation}
The continuity equation $\dot{\rho}+3H(1+w)\rho=0$ 
holds for both $\rho_m$ and $\rho_\phi$, where
$w\equiv p/\rho$ is the equation of state parameter. 
It is interesting to note that, due to the lack of
trivial Einstein frame available through conformal
transformations, the effective equation of state of
 dark energy $w_\phi= p_\phi/\rho_\phi$ 
in teleparallel cosmology can exhibit~\cite{TDE}  
quintessence-like, phantom-like or
phantom-divide crossing evolutions, respectively.

\section{Cosmological perturbations} \label{CP}

\subsection{Perturbation Equations} 

The general vierbein field has 16 components, which are
dynamical  in teleparallel gravity theories. In particular,
6 of the 16 degrees of freedom, arising as a result of the 
non-invariance of
the local Lorentz symmetry, are unfamiliar in the metric perturbation
scenario, and shall not appear in the field equation of the
minimal TEGR limit. The perturbed vierbein field
is given by~\cite{MDP in MTT}
\begin{eqnarray} \label{PV}\nonumber
e^0_\mu &=&\delta^0_\mu(1+\psi)+a\delta^i_\mu\partial_i(F+\alpha)
                               +a\delta^i_\mu(G_i+\alpha_i), \\
e^a_\mu &=&a\delta^a_\mu(1-\varphi)
             +a\delta^i_\mu(\partial_i\partial^a B+\partial^a C_i
             +h^a{}_i)     \nonumber\\
        &~&~~~~~~~~~~~~~~+a\delta^i_\mu B^a{}_i
                         + \delta^0_\mu(\partial^a\alpha+\alpha^a),
\end{eqnarray} 
where $B^a{}_i$ is an antisymmetric spatial tensor with
$B_{ij}+B_{ji}=0$ and $B_{ij}\equiv \delta_i^aB_{aj}$,
while $C_i$, $G_i$, and $\alpha_i$ are transverse vectors
with $\partial^iC_i=\partial^iG_i=\partial^i\alpha_i=0$.
It is easy to obtain the corresponding metric from 
(\ref{PV}) as:
\begin{eqnarray}
g_{00} &=& 1+2\psi \nonumber\\\nonumber
g_{i0} &=& a(\partial_iF + G_i) \\
g_{ij} &=& -a^2[(1-2\varphi)\delta_{ij}+h_{ij}+\partial_i\partial_jB
           +\partial_jC_i+\partial_iC_j], ~~
\end{eqnarray}
which show the familiar metric perturbations in 
GR. Hence, we  observe that the
unfamiliar modes in the perturbed vierbein (\ref{PV}) are
the scalar $\alpha$, the vector $\alpha^i$ and the tensor $B^a{}_i$,
which comprise $1+2+3=6$ degrees of freedom.
Namely, the 16 components of the vierbein field 
are illustrated by the 5 scalar modes:
$\psi$, $\phi$, $B$, $F$ and $\alpha$,  
3 vector modes: $C_i$, $G_i$ and $\alpha_i$,
an antisymmetric tensor $B^a{}_i$, 
and a transverse traceless tensor mode $h_{ij}$, respectively.

In the following study, we focus on the
cosmological implication of the scalar perturbations.
To simplify the calculations, it is convenient to
eliminate the variables  $F$, $B$ and $C_i$ through
the general coordinate transformations
$x^\mu\rightarrow x^\mu+\epsilon^\mu(x)$. 
The vierbein perturbation is provided by~\cite{MDP in MTT}, given by
\begin{eqnarray}\label{scalar mode}
e^0_\mu &=& \delta^0_\mu(1+\psi)+a\delta^i_\mu\partial_i\alpha \nonumber\\
e^a_\mu &=&a\delta^a_\mu(1-\varphi)+a\delta^i_\mu B^a{}_i+ 
            \delta^0_\mu\partial^a\alpha, 
\end{eqnarray}
 leading to the perturbed metric in terms of the
longitudinal gauge form:
\begin{equation} 
\nonumber
ds^2=(1+2\psi)dt^2-a^2(1-2\varphi)\delta_{ij}dx^idx^j.
\end{equation}
Subsqeuently, we obtain
the torsion tensor and  scalar to be
\begin{eqnarray} \nonumber
T^{0}{}_{0i}&=&-\partial_i\psi+ a\partial_0\partial_i\alpha, \\
T^{i}{}_{0j}&=&(H-\dot{\varphi})\delta^i_j+ \partial_0B^i{}_j
               -a^{-1}\partial_j\partial^i\alpha ,  \nonumber\\ 
T^{i}{}_{jk}&=&\partial_k(\delta^i_j\varphi-B^i{}_j)-
               \partial_j(\delta^i_k\varphi-B^i{}_k),
\end{eqnarray}
and 
\begin{equation} 
T=-6H^2+12H(\dot{\varphi}+H\psi)+4a^{-2}\partial^2\alpha_m,
\end{equation}
respectively, where $\alpha_m\equiv aH\alpha$ and
$\partial^2\equiv\delta_{ij}\partial^i\partial^j$.

On the other hand, we can  also 
decompose the scalar field to
homogeneous and perturbed parts:
$\phi\rightarrow \phi(t)+\delta\phi(t,x^i)$.
The scalar perturbations of the matter source 
$\Theta^\mu_\nu$ are denoted by the perfect fluid perturbations:
$\Theta^0_0=-(\rho_m+\delta\rho)$, 
$\Theta^0_i=-(\rho_m+p_m)\partial_i\delta u$ and
$\Theta^i_j=(p_m+\delta p)\delta^i_j$,
where $\delta u$ is the scalar vector potential
of the fluid.

After substituting these expressions to the field equation
(\ref{eom}), we derive the perturbation equations
given  by the energy density $\Theta_0{}^0$:
\begin{eqnarray}
 \label{00}
(1+\kappa^2\xi\phi^2) 
\left[  6H(\dot{\varphi}+H\psi)+2\frac{\partial^2}{a^2}\varphi\right]
+&&\\
 \kappa^2
\left( V_\phi\delta\phi+\dot{\phi}\dot{\delta\phi}-\dot{\phi}^2\psi-6\xi H^2\phi\delta\phi\right) 
&=&
-\kappa^2\delta\rho, 
\nonumber
\end{eqnarray}
the energy flux $\Theta_0{}^i$:
\begin{eqnarray} 
\label{0i}
&& 
2(1+\kappa^2\xi\phi^2)\partial^i(\dot{\varphi}+H\psi)
-\kappa^2 \dot{\phi}\partial^i\delta\phi 
+ 4\kappa^2\xi\phi\dot{\phi}
\left(  \partial^i\varphi+\frac{1}{2}\partial_jB^{ji}\right)
\nonumber\\
&&=-\kappa^2(\rho_m+p_m)\partial^i\delta u, 
\end{eqnarray}
and the momentum density $\Theta_i{}^0$:
\begin{eqnarray} 
\label{i0}
-2(1+\kappa^2\xi\phi^2)\partial_i(\dot{\varphi}+H\psi)
+\kappa^2 \dot{\phi}\partial_i\delta\phi 
+4\kappa^2\xi H\phi\partial_i \delta\phi
=\kappa^2(\rho_m+p_m)\partial_i \delta u,
\end{eqnarray} 
respectively,
while the momentum flux $\Theta_i{}^j$ is divided by
its diagonal components ($i=j$): 
\begin{eqnarray}
 \label{ij}
&&(1+\kappa^2\xi\phi^2) \nonumber
\left[ 6H(\dot{\varphi}+H\psi)+2(\ddot{\varphi}+H\dot{\psi}+2\dot{H}\psi)\right]
-\kappa^2\xi\phi\left( 2\dot{H}+3H^2\right)\delta\phi  
\nonumber
\\
&&-\kappa^2\left[ \dot{\phi}\delta\dot{\phi}-V_\phi\delta\phi-\dot{\phi}^2\psi\right] 
 +4\kappa^2\xi\phi\dot{\phi}(\dot{\varphi}+2H\psi) 
         - 4\kappa^2\xi H(\dot{\phi}\delta\phi+\phi\dot{\delta\phi})
          =\kappa^2\delta p\,,
\end{eqnarray}
as well as its anisotropic parts ($i\neq j$):
\begin{equation} 
\label{inj}
\psi=\varphi - 
\frac{2\kappa^2\xi\phi\dot{\phi}}{H(1+\kappa^2\xi\phi^2)}\alpha_m.
\end{equation}
Similarly, the perturbed equation of motion for the scalar field
(\ref{eom phi}) is given by
\begin{eqnarray}
 \label{eom phi2}
&&\ddot{\delta\phi}
+3H\dot{\delta\phi}
+\left( V_{\phi\phi}-\frac{\partial^2}{a^2}\right) \delta\phi
+6\xi H^2 \delta\phi 
-2(\ddot{\phi}+3H\dot{\phi})\psi
-\dot{\phi}(3\dot{\varphi}+\dot{\psi}) \nonumber\\
&&
-12\xi H\phi\left( \dot{\varphi}+H\psi\right) 
-4\xi \phi\frac{\partial^2}{a^2}\alpha_m =0.
\end{eqnarray}

To complete the perturbation equations 
of teleparallel dark energy models,
we shall also include equations from the constraint 
(\ref{constraint}). In linear perturbations,
this constraint yields
\begin{equation} \label{c2}
\dot{\phi}\partial^i\varphi
+\frac{1}{2}\dot{\phi}\partial_jB^{ji}
+H \partial^i \delta \phi=0,
\end{equation}
which is necessary for the consistent relation
of the matter source $\Theta_{0i}=\Theta_{i0}$,
as also indicated from the comparison between
(\ref{0i}) and (\ref{i0}).
Given that the antisymmetric tensor $B_{ij}$
satisfies $\partial_i\partial_jB^{ij}=0$,
we find that the gradient of
(\ref{c2}) leads to
\begin{equation} \label{c3}
\varphi=-\frac{H}{\dot{\phi}}\delta\phi.
\end{equation}
Note that this equation holds provided that
the non-minimal coupling $\xi\neq 0$, while
 the antisymmetric tensor $B_{ij}$
does not couple to other variables.
{\footnote{It has been studied in a recent
work~\cite{revisit} 
that $B_{ij}$ can be attributed to a pseudo-scalar mode
and a pseudo-vector mode, which have no dynamical importance in the
linear perturbation of $f(T)$ gravity.}}

\subsection{Numerical results}

Although teleparallel gravity gives
a gravitational concept, which is different from GR,
the perturbations of the matter sector shall be
treated in the usual manner regardless to
the geometrical background. In particular, 
the equation motion of a particle in the
presence of gravity in TEGR is mathematically
equivalent to the geodesic equation in GR 
\cite{Maluf:1994ji},
while the conservation of the energy-momentum
$\nabla^\mu \Theta_{\mu\nu}=0$ is the covariant
derivative with respect to the GR connection
\cite{LV,Weinberg}.
Thus, the density perturbation of the dust-like source,
$\delta_m \equiv\delta\rho_m/\rho_m$, still satisfies
the perturbation
\begin{equation} \label{deltam}
\ddot{\delta}_m+2H\dot{\delta}_m=
\frac{\partial^2}{a^2}\psi+
3\ddot{\varphi}+6H\dot{\varphi},
\end{equation}
where we have used $p_m=\delta p=0$. The gauge invariant
definition $\Delta\equiv\delta_m+3H\delta u$ is also
applied to the new variable 
$\Phi\equiv\varphi+H\delta u$, and the matter
perturbation now takes the form
\begin{equation}
\ddot{\Delta}+2H\dot{\Delta}=
\frac{\partial^2}{a^2}\psi+
3\ddot{\Phi}+6H\dot{\Phi}.
\end{equation}

For the non-vanished coupling $\xi\neq 0$, the constraint
(\ref{c3}) holds, so that we can replace the perturbation of the
scalar field $\delta\phi$ by the gravitational potential
$\varphi$. Since $\alpha_m$ is described by the difference
between $\psi$ and $\varphi$, it is straightforward to
rewrite the perturbation equations in terms of
the gravitational potentials  similar to GR. 
As a result, we may use, for instance, (\ref{ij}) 
and (\ref{eom phi2}) to solve the evolutions of
$\psi$ and $\varphi$, and obtain the result of the
density perturbation $\delta_m$ through (\ref{00}).
In particular,
during the matter dominated epoch where both the scalar field
variables $\phi$ and $\delta\phi$ are negligible, the
equations (\ref{00})-(\ref{inj}) can be reduced to the same forms
as those in GR. We can drop all the gradient terms by considering
the time earlier than the perturbation modes entering the horizon,
leading to the initial conditions of the
gravitational potentials as $\psi=\varphi=constant$, while
 (\ref{00}) implies $\delta_m(z\gg 1)=-2\varphi$.

\begin{figure}[!]
\begin{center}
\includegraphics[width=7.5cm]{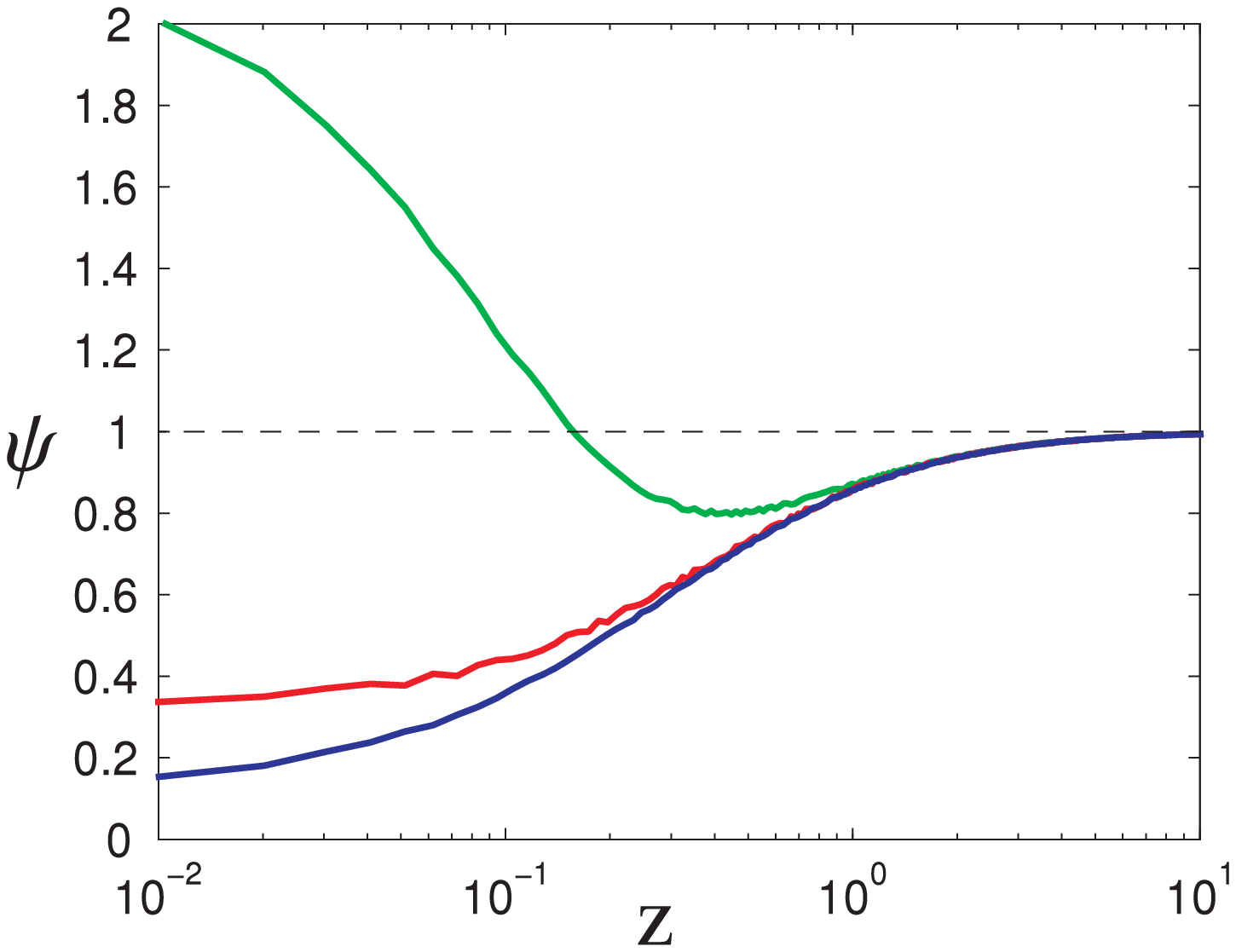}
\includegraphics[width=7.5 cm]{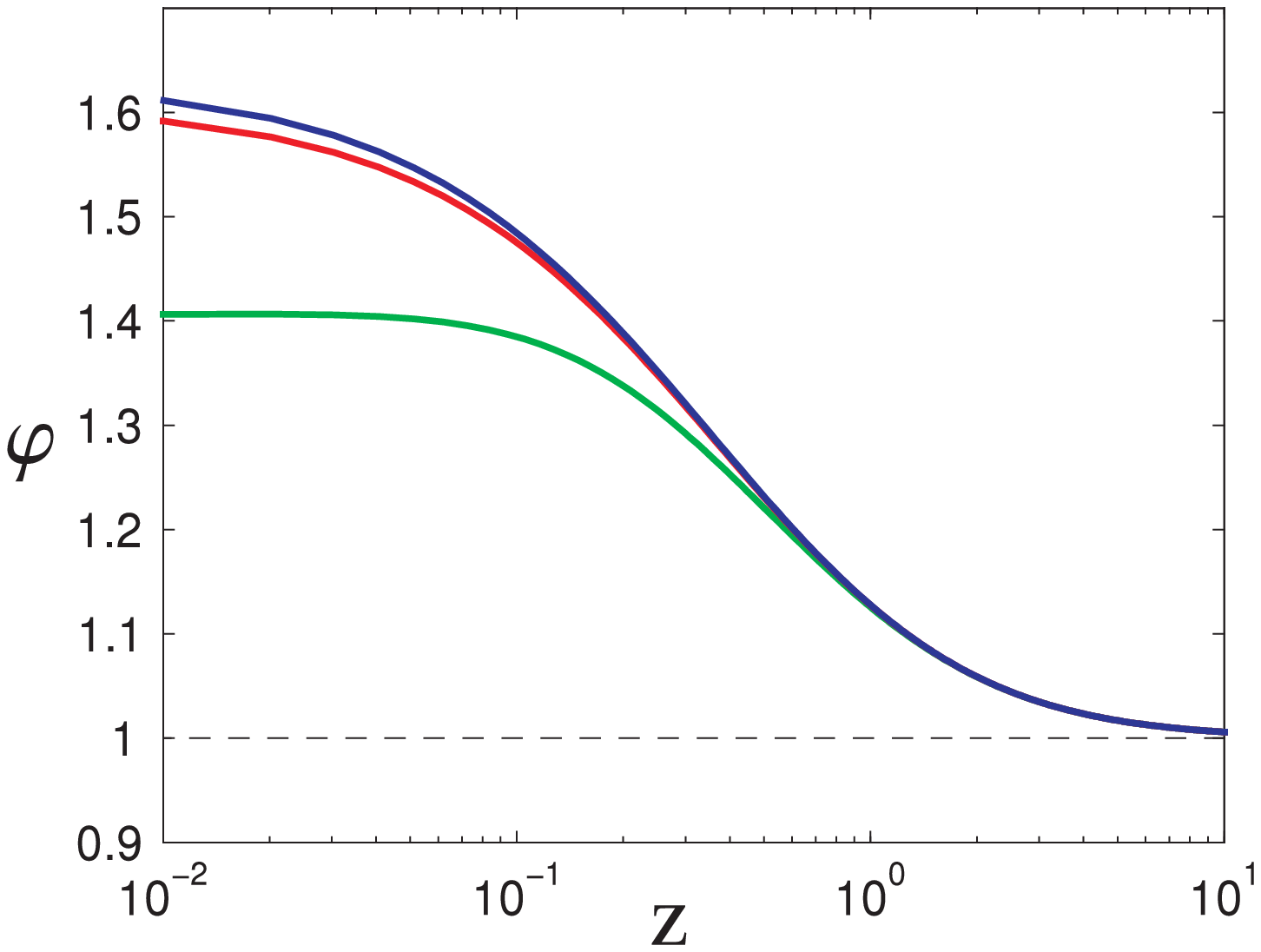}
\caption{Evolutions of the gravitational potentials
$\psi$ (left panel) and $\varphi$ (right panel) with
respect to the redshift $z$ for the potential-less ($V=0$) 
teleparallel dark energy model with $\xi=-0.343$,
where the green, red and
blue solid lines correspond to the wavenumber
$k=3\times 10^{-4}\,,\ 1\times 10^{-3}$
and $1\times 10^{-2}\,h\;\mathrm{Mpc}^{-1}$, respectively,
and the initial value is normalized to unity for $z\gg 1$.}
\label{psiphi}         
\end{center}
\end{figure}

\begin{figure}[!]
\begin{center}
\includegraphics[width=7.5cm]{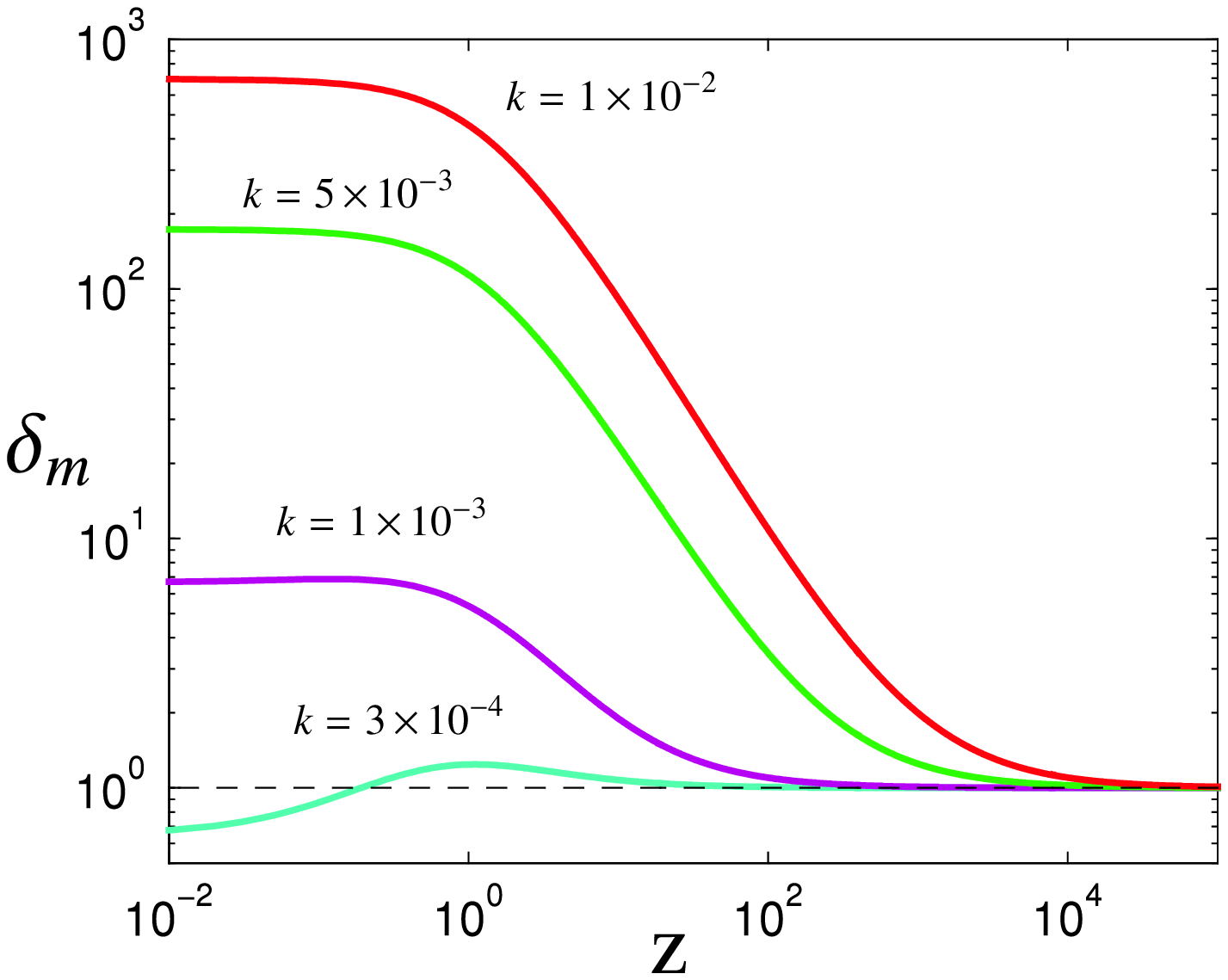}
\includegraphics[width=7.5 cm]{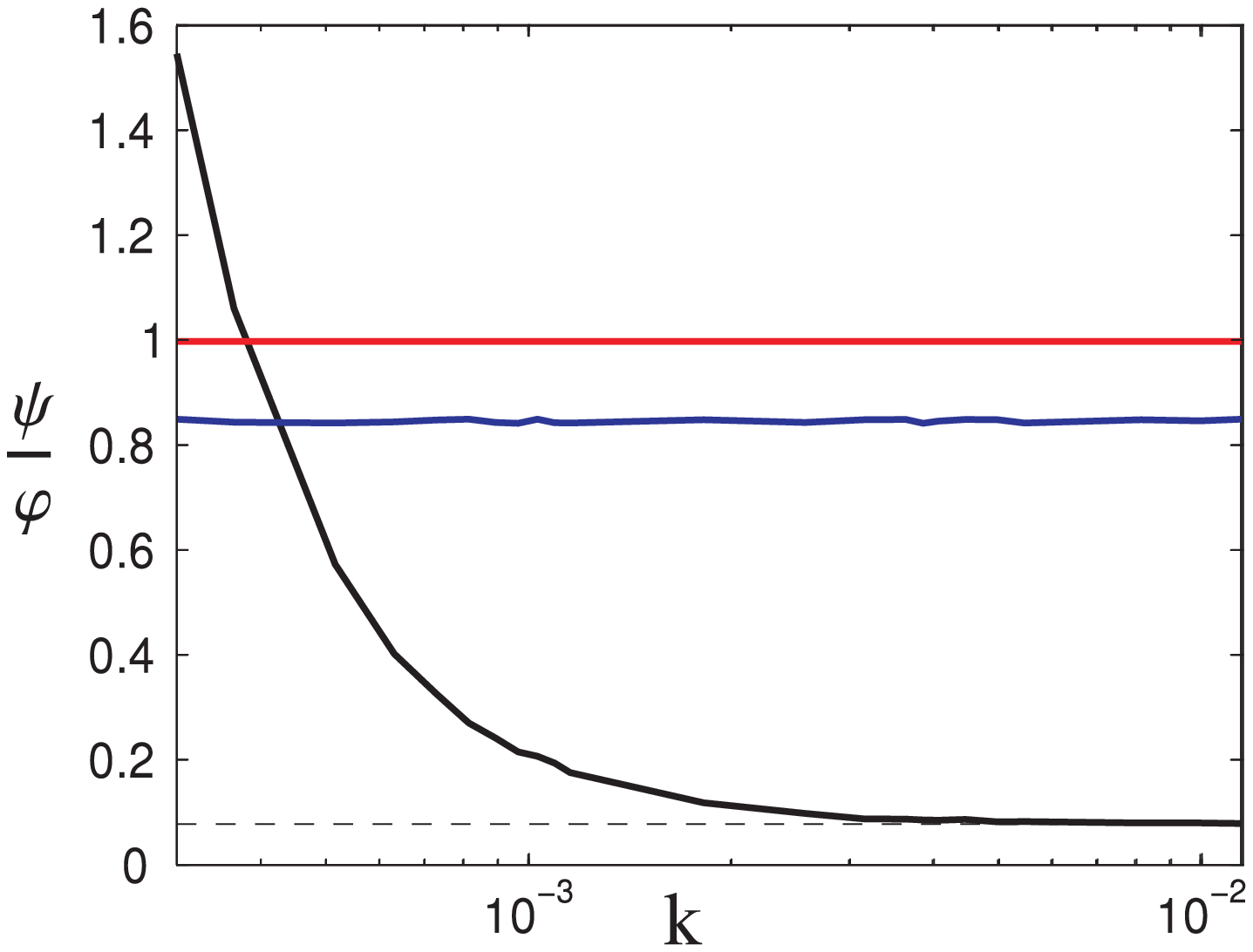}
\caption{
Evolutions of 
the matter density perturbation $\delta_m$ (left panel)
with respect to the redshift $z$ for various of wavenumbers 
$k$ ($h\;\mathrm{Mpc}^{-1}$)
with the initial value is normalized to unity for $z\gg 1$
and the ratio $\psi/\varphi$ (right panel)
with respect to the wavenumber $k$ at present ($z=0$)
in the potential-less ($V=0$) 
teleparallel dark energy model.
In the right panel,
the dashed line  corresponds to $\psi/\varphi= 0.078$ 
given by  (\ref{4.4}) in 
the sub-horizon limit ($k\gg a_0H_0$), while
the black, blue and red solid ones represent 
 different non-minimal couplings of 
$\xi=-0.343$, $-0.125$ and $-0.05$, respectively.}
\label{deltam to z}         
\end{center}
\end{figure}

We solve the evolutions of the perturbed variables
numerically as seen in 
Figs.~\ref{psiphi} and  \ref{deltam to z}.
Here, we concentrate on the potential-less teleparallel
dark energy model ($V=0$) with initial values
$\kappa\phi_i=4.03\times 10^{-8}$ and
$\kappa\dot{\phi}_i/H_i=3.5\times 10^{-8}$, 
where we require the current matter density 
$\Omega_{m0}=0.3$.
It is evident that $\psi$ and $\varphi$ 
 behave differently when $z\lesssim 10$, in 
which the energy density of
$\phi$ becomes important.
In the right panel of
Fig. \ref{deltam to z}, we illustrate this difference
at present and we find that the value of $\psi$ is 
greater than $\varphi$ on super-horizon scales but
decreases with the scales. The ratio $\psi/\varphi$
approaches to 0.078 on the small scales, which is
consistent to the result (\ref{4.4}) 
obtained from the quasi-static approximation in the
sub-horizon regime.
We note that in the right panel of Fig.~\ref{deltam to z}, we have 
shown  $\psi/\varphi$ for different values of the non-minimal 
coupling $\xi$. It is easy to see that it approaches to the GR limit for
the small $\xi$ as expected.

The evolutions of the gravitational potentials with respect 
to time suggest a non-vanishing part on the right hand side 
of (\ref{deltam}) on super-horizon sacles, 
which generates deviations
from the $\Lambda$CDM model. In the left panel of 
Fig. \ref{deltam to z}, we find that the effect becomes
significant around the horizon scale of
$k\sim 10^{-4}h\, \mathrm{Mpc}^{-1}$. Consequently,  the
imprint of the non-dynamical new degree of freedom
$\alpha_m$ could be strictly challenged by observations
on the large scale structure of the potential-less 
teleparallel dark energy model.

\section{Matter Growth} \label{MG}

In this section, we study the cosmological perturbations
on sub-horizon scales $k\gg aH$, where the density perturbation
obeys the modified equation from GR
\begin{equation} \label{PE}
\ddot{\delta}_m+2H\dot{\delta}_m-4\pi G_{\mathrm{eff}}\rho_m\delta_m = 0.
\end{equation}
The effective gravitational coupling constant $G_{\mathrm{eff}}$
is expected to coincide with the Newton's constant 
in the minimal case when $\xi=0$.

We proceed a further simplification by
adopting the quasi-static approximation
\begin{equation}
 \nonumber
\vert\dot{X}\vert\lesssim \vert HX\vert,
\;\;\;\;\;\mbox{for}\; X=\phi,\psi,\varphi,
\end{equation}
to the perturbation equations 
(\ref{00})-(\ref{eom phi2}).
Hence, the effective gravitational coupling
can be obtained from the modified Poisson equation
$4\pi G_{\mathrm{eff}}\rho_m\delta_m\simeq k^2\psi/a^2$. 
Under this approximation, we have $\Delta\simeq\delta_m$,
and together with the sub-horizon condition
$k\gg aH$, we find that 
(\ref{eom phi2}) becomes
\begin{equation} \label{sub phi}
\left( \frac{k^2}{a^2}+V_{\phi\phi}\right) \delta\phi
+4\xi \phi\frac{k^2}{a^2}\alpha_m \simeq 0,
\end{equation}
while  (\ref{00}) gives
\begin{equation} \label{sub 00}
-2(1+\kappa^2\xi\phi^2)\frac{k^2}{a^2}\varphi
+\kappa^2V_\phi\delta\phi\simeq-\kappa^2\delta\rho,
\end{equation}
where $\vert\xi\vert$ is 
 $O(10^{-1})$ according to
 the observational  constraints~\cite{ob TDE}.
We shall further neglect the contributions
of $V_\phi$ and $V_{\phi\phi}$ as for the general
scalar-tensor theory of dark energy~\cite{scalar-tensor}.
It can be checked that any potential for teleparallel
dark energy indeed satisfies 
$k^2\gg a^2\kappa V_\phi$ and
$k^2\gg a^2V_{\phi\phi}$ 
in the sub-horizon regime \cite{ob TDE}.

 From (\ref{c3}) and (\ref{sub phi}), we obtain that
$\delta\phi=-\dot{\phi}\varphi/H
\simeq 4\xi \phi\alpha_m$. The relation (\ref{inj}) results in
\begin{equation}
\label{4.4}
\psi=\left( 1-\frac{\epsilon}{1+\kappa^2\xi\phi^2}\right)\varphi, 
\end{equation}
where  $\epsilon\equiv\kappa^2\dot{\phi}^2/2H^2$.
From (\ref{sub 00}) and (\ref{4.4}), we finally arrive at
\begin{equation} 
\frac{k^2}{a^2}\psi=\frac{1}{1+\kappa^2\xi\phi^2}
\left( 1-\frac{\epsilon}{1+\kappa^2\xi\phi^2}\right)
\frac{\kappa^2}{2}\delta\rho,
\end{equation}
which implies 
\begin{equation} \label{G eff}
G_{\mathrm{eff}}=
\left( 1-\frac{\epsilon}{1+\kappa^2\xi\phi^2}\right)
\frac{G}{1+\kappa^2\xi\phi^2}.
\end{equation}
Note that $\epsilon$ is proportional to the 
kinetic energy density of the scalar field, 
which is sub-dominated to the energy density
of dark energy $\Omega_\phi$ at the present
in teleparallel dark energy models.
In the minimal quintessence limit $\xi=0$, 
the cosmic acceleration in fact requires
$\epsilon\ll \Omega_\phi$ and the effective
gravitational coupling reduces to the expected
result, $i.e.$
$G_{\mathrm{eff}}=(1-\epsilon)G\simeq G$.
 
\subsection{Purely gravity coupling $V=0$}

In the following, we exam the matter growth in
teleparallel dark energy models with specific 
forms of potentials.
The simplest tracker behavior has been found in the 
potential-less case ($V=0$) where one can achieve
the late-time cosmic acceleration for $\xi<0$,
 while the observational constraints from
type Ia supernova (SNIa), baryon acoustic oscillation (BAO)
and cosmic microwave background (CMB) data reveal a 
theoretically viable region of $-1<\xi<-0.125$ with
the best-fit value at $\xi=-0.35$ \cite{Gu}. 
To analyze the evolution in this case, 
we re-parametrize two new variables:
$x=\kappa\dot{\phi}/\sqrt{6}H$ and
$y=\sqrt{-\xi}\kappa\phi$ so that
the Friedmann equation (\ref{FR1}) is rewritten as
\begin{equation} 
1=\Omega_m+x^2+y^2.
\end{equation}
The evolutions of $x$ and $y$ with respect to 
$N=\ln a$ are given by
\begin{eqnarray} 
\nonumber
\frac{dx}{dN}&=&-\left( 3+\frac{\dot{H}}{H^2}\right)x
+\sqrt{-6\xi}y,  \\
\frac{dy}{dN}&=& \sqrt{-6\xi}x.
\end{eqnarray}
Note that  (\ref{FR2}) also provides a useful relation:
\begin{equation} \label{H dot}
(1-y^2)\frac{\dot{H}}{H^2}=
-3x^2+2\sqrt{-6\xi}xy-\frac{3}{2}\Omega_m,
\end{equation}
to solve the evolutions
numerically from some given initial conditions.

\begin{figure}[!]
\begin{center}
\includegraphics[width=6.5cm]{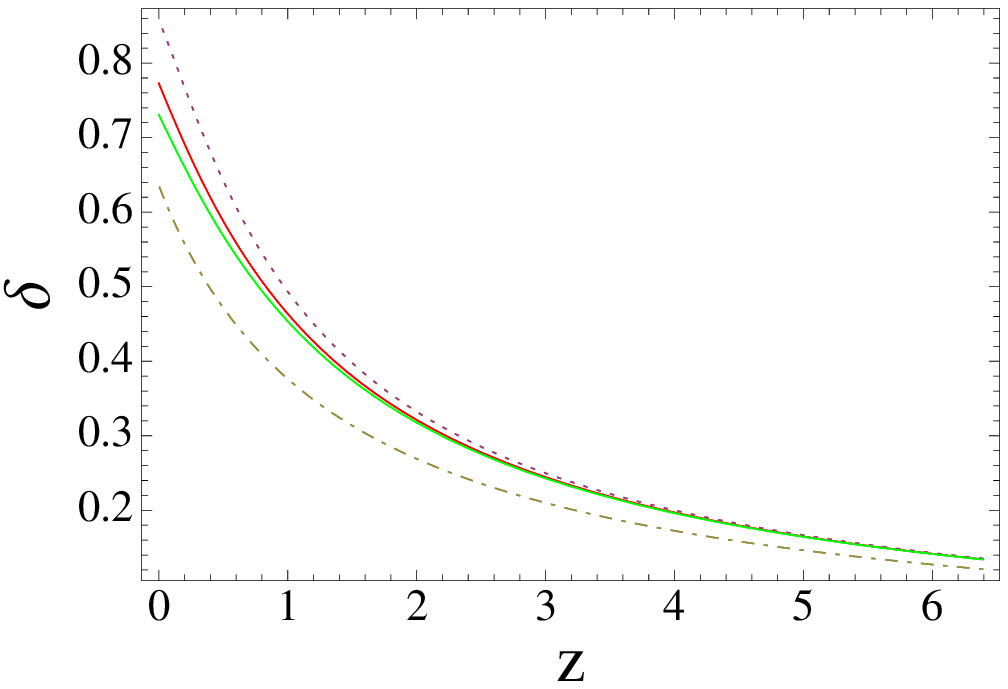}
\;
\includegraphics[width=6.5 cm]{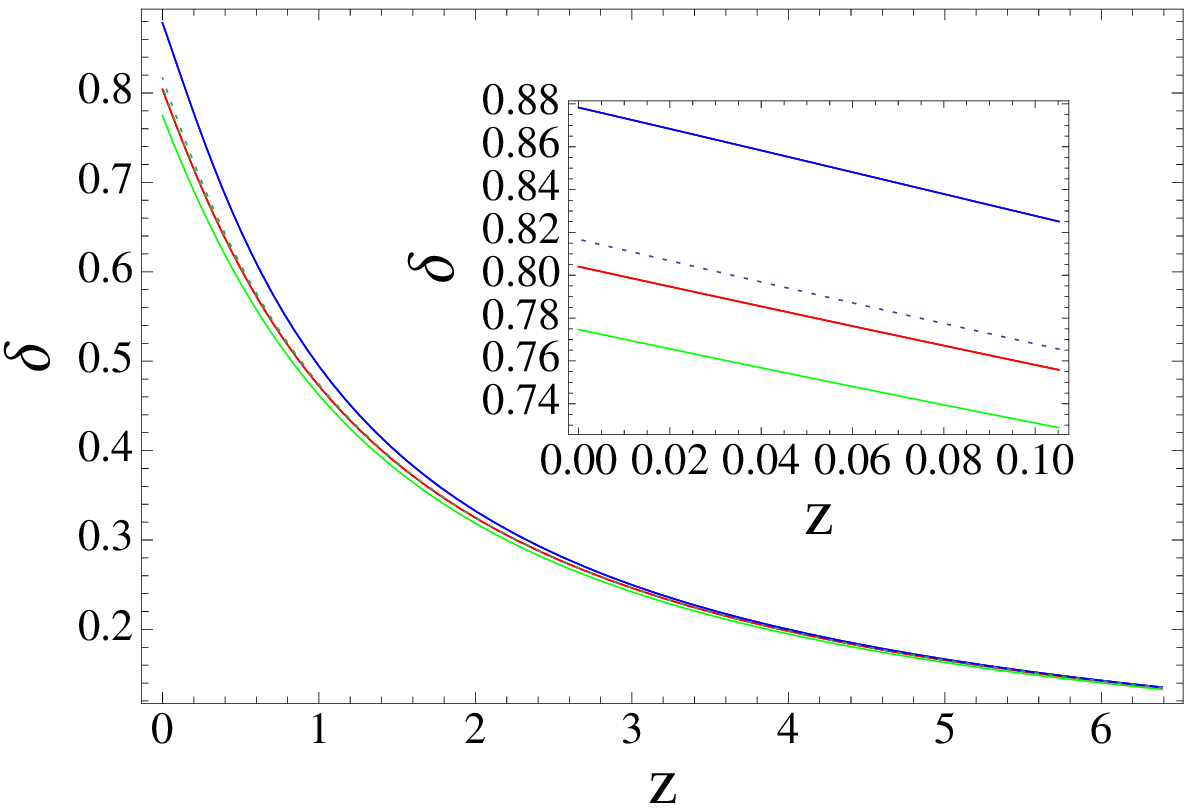}
\caption{
Growth rate $\delta$ as a function of the redshift $z$ in the 
teleparallel dark energy model with $V=0$ ($V=V_0\,e^{-2\kappa\lambda\phi}$)
for the left (right) panel.
In the left figure, the dotted,  red-solid and 
 dot-dashed lines correspond to the lower,
 best-fit and upper values of 
$\xi=-1$, $-0.35$ and $-0.125$, respectively,
while the green-solid line represents $\xi=-0.35$ with 
a fixed value of $Q=1$.
%
In the right one, the red-solid,  
blue-solid and green-solid lines represent  
$\xi=-0.4$, $-1$ and $-0.125$ with the initial values of 
$\left\lbrace \Omega_m=0.999\,,\,v= 1.0\,\times 10^{-6}\right\rbrace$,
and the dotted line depicts $\xi=-0.4$ with the initial values
$\left\lbrace \Omega_m=0.990\,,\,v= 6.0\,\times 10^{-3}\right\rbrace$,
while 
$\lambda=0.5$ and the initial values 
$\left\lbrace y= 1.0\times 10^{-6}\,
         ,\,\mathcal{G}=0\right\rbrace $ for $z\gg 1$ are fixed.}
\label{exp}         
\end{center}
\end{figure}

By using $\mathcal{G}\equiv d\ln(\delta/a)/d\ln a$
to re-parametrize the matter perturbations,
the second derivative equation
(\ref{PE}) is reduced to
\begin{equation} \label{G1}
\frac{d~\mathcal{G}}{d N}
+\left( 2+\frac{\dot{H}}{H^2}\right)(\mathcal{G}+1)+(\mathcal{G}+1)^2
= \frac{3}{2}Q\,\Omega_m\,,
\end{equation}
where $Q\equiv G_{\mathrm{eff}}/G$. Subsequently,
 from (\ref{G eff}) we find that 
\begin{equation}
Q=\frac{1}{1-y^2}\left( 1-\frac{\sqrt{6}x^2}{2(1-y^2)}\right).
\end{equation}
We remark that (\ref{G1}) is applicable to
the generic modified gravity theories 
if the effective coupling $G_{\mathrm{eff}}$ is given.
In the GR case, the contribution of the
$\mathcal{G}^2$ term in (\ref{G1})
is negligible throughout the growth history
from matter domination~\cite{Linder APS}.
Here, we adopt the same approximation 
given that the present value of
$\mathcal{G}$ is not smaller than $-0.5$ 
for all of the considered cases in this work.
As a result, we can have
the evolution of $\mathcal{G}$
 solved simultaneously with the other parameters
$x$ and $y$ from given initial conditions.
In the left panel of Fig.~\ref{exp}, we show the evolution of the
growth rate 
$\delta$ as a function of the redshift $z$ in the purely gravity
coupling teleparallel dark energy model ($V=0$),
where the dotted,  red-solid and 
 dot-dashed lines correspond to the observational lower bound,
 best-fit and upper bound values of 
$\xi=-1$, $-0.35$ and $-0.125$ respectively,
while the green-solid line represents $\xi=-0.35$ with 
a fixed value of $Q=1$ so that
gravitational interaction mimics that of GR.
The initial values provided are
$\left\lbrace\Omega_m=0.999\,,\, y= 1.0\times 10^{-6}\,
         ,\,\mathcal{G}=0\right\rbrace $    
for $z\gg 1$.
The results in the left panel of Fig.~\ref{exp} indicate that
the gravitational interaction is in fact
enhanced ($Q>1$) during the evolution, which ensures
a more reasonable approximation by
dropping the $\mathcal{G}^2$ term in (\ref{G1}).

\subsection{Exponential potential $V=V_0\,e^{-2\kappa\lambda\phi}$}

We can extend the scenario to teleparallel dark energy
models with non-vanished potentials. 
As an illustration, we concentrate on the exponential type,
$V=V_0\,e^{-2\kappa\lambda\phi}$, which has a
constant potential limit
$V=V_0$ if $\lambda=0$.
We shall introduce a new parameter,
$v\equiv \kappa\sqrt{V}/\sqrt{3}H$, to address the
evolution of the potential. The
 Friedmann equation (\ref{FR1}) is rewritten as
\begin{equation} 
1=\Omega_m+x^2+y^2+v^2,
\end{equation}
where $x$ and $y$ share the same definitions as
 the previous case.
The evolution equations of these components are 
given by
\begin{eqnarray}\nonumber
\frac{dx}{dN}&=&-\left( 3+\frac{\dot{H}}{H^2}\right)x
+\sqrt{-6\xi}y+\sqrt{6}\lambda v^2,  \\ \nonumber
\frac{dy}{dN}&=& \sqrt{-6\xi}x,    \\
\frac{dv}{dN}&=&-\left(\sqrt{6}\lambda x+\frac{\dot{H}}{H^2}\right)v.
\end{eqnarray}
We consider  the regime of
$\xi<0$ for the exponential potential 
since it is favored from SNIa, BAO and
CMB observations \cite{ob TDE}
and solve $x$, $y$ and $v$ together
with the density perturbation
$\mathcal{G}= d\ln(\delta/a)/d\ln a$
in (\ref{G1}).
Note that (\ref{H dot}) is unchanged.

In the case with the non-minimal coupling
of  $\xi\lesssim -0.01$, the 
result
(refer to the green-solid line in the right panel of Fig.~\ref{exp})
is similar to the quintessence model
where the potential dominates during
the evolution. 
However, if $\xi$ is around 
the best-fit value of $\sim -0.4$~\cite{ob TDE}, 
the purely gravity coupling
tracker behavior appears when the given
initial value of $v$ is sufficiently small,
as illustrated by the dotted line
in the right panel of Fig.~\ref{exp} with the initial value
$v\sim 10^{-6}$. The comparison to the red-solid
line shows that the gravitational interaction
is weaker if the cosmic acceleration is driven
by the potential instead of the NGF tracker effect.
On the other hand, if the non-minimal coupling $\xi$
is around $-1$ as described
by the blue-solid line in the right panel of Fig.~\ref{exp}, 
then only NGF
tracker behavior is found regardless to the
initial values of the potential. Clearly,
the parameter $y$ will finally dominate
the dark energy density $\Omega_\phi$ at the
present.

\section{Conclusions} 

In this work, we have applied the vierbein 
perturbation scenario to study the cosmological
perturbations in teleparallel dark energy models
with the non-minimal coupling between the scalar field and gravity.
Unfamiliar with the metric perturbations, a new scalar mode, 
$\alpha_m$, appears in the linear equations and mainly
contributes to the anisotropic between the
gravitational potentials $\psi$ and $\varphi$.
This new scalar degree of freedom is fully suppressed 
in the sub-horizon scales of $f(T)$ gravity \cite{MDP in MTT},
but has significant contributions in the 
large scale structures \cite{LSS}.
Nevertheless, the dynamical degrees of freedom
of the theory do not increased by $\alpha_m$
given that the constraint in (\ref{constraint})
provides an additional algebraic relation to the 
scalar mode variables.
We have demonstrated 
the numerical results for  the effects
of $\alpha_m$ in terms of the ratio $\psi/\varphi$,
and  found that the discrepancy from GR is significant
from sub-horizon scales well up to super-horizon
scales.

To study the matter growth in teleparallel cosmology,
we have used the quasi-static approximation in the
sub-horizon scales. The effective gravitational coupling
becomes scale independent in the sub-horizon
regime similar to the generic scalar-tensor theories.
In the case where the cosmic acceleration is driven by
the non-minimal gravity-field coupling term, i.e.
the $\xi H^2\phi^2$ term in  (\ref{telerho}), we have
found that the gravitational interaction is enhanced
so that matter grows faster than that in GR.
In the broader case where the scalar field has a
potential, the quintessence behavior recovers
with a small coupling limit of $\xi$, while
the non-minimal gravity-field tracker is followed
when $\xi\sim-1$. 
In the case with $\xi$ around the best-fit
value ($\sim -0.4$) from observations, 
we can also observe a faster growth rate 
if the NGF tracker is followed instead of the usual 
quintessence-like evolution. 
These results feature the matter perturbations in
teleparallel dark energy models and worth for further 
investigations to look for the possible deviations from GR.


\section{Acknowledgements}
We are grateful to helpful comments given by 
Qing-Guo Huang and Eric Linder as well as
Chung-Chi Lee for his technical support. We
 thank KITPC for the wonderful program 
``Cosmology and Astroparticle Physics.''
The work was supported in part by National Center of Theoretical Science
and  National Science Council 
(NSC-98-2112-M-007-008-MY3)
of R.O.C.

\end{document}